\documentclass[twocolumn,secnumarabic,amssymb, nobibnotes, aps, prd, showpacs,amsmath]{revtex4}
\usepackage[dvipdf]{graphicx}

\usepackage{graphicx}
\usepackage{dcolumn}
\usepackage{bm}
\usepackage{color}

\begin{document}

\title{General mechanism for amplitude death in coupled systems}
\author{V. Resmi}
\email{v.resmi@iiserpune.ac.in}
\affiliation{Indian Institute of Science Education and Research, 
Pune - 411021, India} 
\author{G. Ambika}
\email{g.ambika@iiserpune.ac.in}
\affiliation{Indian Institute of Science Education and Research, 
Pune - 411021, India} 
\author{R. E. Amritkar}
\email{amritkar@prl.res.in}
\affiliation{Physical Research Laboratory, Ahmedabad - 380009, India}
\begin{abstract}
  We introduce a general mechanism for amplitude death in coupled synchronizable dynamical systems. It is known that when two systems are coupled directly, they can synchronize under suitable conditions. When an indirect feedback coupling through an environment or an external system is introduced in them, it is found to induce a tendency for anti-synchronization. We show that, for sufficient strengths, these two competing effects can lead to amplitude death. We provide a general stability analysis that gives the threshold values for onset of amplitude death. We study in detail the nature of the transition to death in several specific cases and find that the transitions can be of two types - continuous and discontinuous. By choosing a variety of dynamics for example, periodic, chaotic, hyper chaotic, and time-delay systems, we illustrate that this mechanism is quite general and works for different types of direct coupling, such as diffusive, replacement, and synaptic couplings and for different damped dynamics of the environment.
\end{abstract}

\maketitle

\section{INTRODUCTION}
The dynamics of many complex systems can be understood as the collective behavior of a large number of dynamical units coupled via their mutual interactions. The dynamics of such connected systems has been an interesting topic of study especially due to its relevance in understanding a large variety of natural systems. Based on the the nature of interactions among the coupled units, they can exhibit many emergent phenomena such as synchronization, hysteresis, phase locking, amplitude death, and oscillator death \cite{Pikovsky2003,Kaneko1993,Ott1993}. Among these, the phenomenon of synchronization is the most widely studied and has relevance in many contexts, such as neuronal networks, communication, laser systems, etc. \cite{Pikovsky2003}. So also, the quenching or suppression of dynamics called amplitude death is another emergent phenomenon of equal relevance in such systems. This can lead to interesting self-adjustable control mechanisms and plays a prominent role as an efficient regulator of the dynamics. The occurrence of amplitude death has been reported in many cases, such as chemical reactions \cite{Dol88,Cro89,Bar85,Dol96}, biological oscillators \cite{Ata06, Kos10, Ozd04}, coupled laser systems \cite{Her00,Wei07} and relativistic magnetrons \cite{Ben89}. We would like to project the importance of the phenomenon of amplitude death in coupled systems in two contexts: one, as a desirable control mechanism in cases such as coupled lasers where it leads to stabilization \cite{Kim05,Kum08} and two, as a pathological case of oscillation suppression or disruption in cases like neuronal disorders such as Alzheimer's disease, Parkinson's disease, etc. \cite{Sel00,Tan05,Cau03}.

The mechanisms so far reported to induce amplitude death in coupled systems are de-tuning of oscillators under strong coupling \cite{Mir90,Erm90,Zhai04}, coupling through conjugate variables \cite{Kar07,Das10}, dynamic coupling \cite{Kon03}, and delay in coupling due to finite propagation or information processing speeds \cite{Ram98,Pra05,Choe07,Dod04}. Distributed delays rather than discrete or constant delays have been proposed as more realistic models in ecology and neurobiology, where the variance of the delay plays a relevant role \cite{Atay03}. So also, amplitude death has been studied in the context of attractive and repulsive couplings in two chaotic Lorenz systems \cite{Chen09}. In all these mechanisms, death occurs dynamically due to the targeting of the units to one or more of the equilibrium states or due to the stabilization of one of these states. The equilibrium states or fixed points can be either that of the uncoupled system or those evolved by coupling. While these mechanisms can model the amplitude death observed in coupled systems of oscillators, we find that all these methods are system specific and may not work in a general case. In the case of death by delay coupling, the limitations of the method have been reported in several cases such as periodic \cite{Kon03} and chaotic systems \cite{Kon04, Kon05}. Moreover, there are many cases such as neuronal disorders where depression of activity or death is due to the presence of another agency or medium \cite{Sel00,Tan05}. For such cases, the mechanism of death is still not fully understood and none of the above mentioned mechanisms so far reported is applicable.

In this work, we introduce a mechanism for amplitude death caused by an indirect feedback coupling through a dynamic environment, in addition to direct coupling.We essentially project the role of the environment in controlling the dynamics of connected systems. We find that, while it essentially explains quenching of activity or suppression induced by an external medium or agent, this method can also serve as a general mechanism to induce death in coupled synchronizable systems. Its generality lies in the fact that it seems to work in any coupled system that can synchronize. It is effective in quenching dynamics in a variety of systems such as periodic oscillators, chaotic systems, hyperchaotic systems, and delay systems. We show that this method also induces amplitude death in systems with different forms of direct coupling interactions, like diffusive, replacement, synaptic coupling, etc. Specifically, we demonstrate that the present method, with a varied model for the environment, works in the case of hyperchaotic systems for which delay coupling is not effective to induce amplitude death. As such, it is an important step in methodology toward achieving controls or stabilization to desirable performance in many practical cases.  The relevance of this method lies in the fact that death can be engineered and can be easily implemented in any system with coupled synchronizable units.

In the present work we use the indirect feedback coupling through the environment of our earlier work where we showed that such a coupling can induce anti-phase- (or anti) synchronization in two systems which are not directly connected \cite{Res10}. Consider two systems coupled directly such that with adequate strength of coupling they can exhibit synchronous behavior. Then if we introduce an additional indirect feedback coupling through the environment or another external system such that it induces a tendency for anti-synchronization, then for sufficient strengths, these two competing tendencies can lead to amplitude death. We find that, in the state of amplitude death, the subsystems stabilize to a fixed point of the coupled system. We also show that the method introduced here can induce amplitude death in coupled systems with different types of dynamics for the environment and for different types of direct coupling.

We develop an approximate stability analysis which provides the threshold or critical values of the coupling strength for amplitude death in the general context. Direct numerical simulation giving the regions of amplitude death in the space of coupling strengths agrees well with the transition curves obtained from the stability analysis.

We also analyze in detail the nature of the transition to the amplitude death state.We find that all the specific cases studied exhibit  either continuous or discontinuous transitions to death.  In the continuous case, as illustrated by two coupled R\"ossler systems, during the transition the full reverse period-doubling scenario is observed, and the system reaches a one-cycle state before amplitude death occurs. The transition to death then occurs due to a super critical Hopf bifurcation. In the discontinuous case, the transition is sudden due to the disappearance of a distant attractor and stabilization of a fixed point. For two coupled Lorenz systems, we find that the transition to death is probably via a sub-critical Hopf bifurcation with long transients, and prior to this, the systems go through a state of frustration between synchronized and anti-synchronized behavior.
\section{Amplitude death via direct and indirect coupling}
We start with two systems coupled mutually with two types of coupling, namely a direct diffusive coupling and an indirect coupling through an environment. The dynamics can be written as
\begin{subequations}
\label{eq:model}
\begin{eqnarray}
\label{eq:model1}
\dot{x}_1 &=& f(x_1) + \varepsilon_d \beta (x_2-x_1) + \varepsilon_e \gamma y,  \\
\label{eq:model2}
\dot{x}_2 &=& f(x_2) + \varepsilon_d \beta (x_1-x_2) + \varepsilon_e \gamma y,  \\
\label{eq:model3}
\dot{y} &=& -\kappa y - \frac{\varepsilon_e}{2} \gamma^{T} (x_1+x_2).
\end{eqnarray}
\end{subequations}
Here, $x_1$ and $x_2$ represent two $m$-dimensional oscillators whose intrinsic dynamics is given by $f(x_1)$ and $f(x_2)$ respectively. The systems are mutually coupled using diffusive coupling [ the second term in Eqs.~(\ref{eq:model1}] and~(\ref{eq:model2})). The environment is modeled by a one-dimensional overdamped oscillator $y$ with a damping parameter $\kappa$. The environment is kept active by feedback from both systems as given by the last term in Eq.~(\ref{eq:model3}). Both systems also get feedback from $y$ [ the last term in Eqs.~(\ref{eq:model1} and \ref{eq:model2})]. $\beta$ is a matrix ($m \times m$) with elements $0$ and $1$ and defines the components of $x_1$ and $x_2$ that take part in the diffusive coupling. For simplicity, we take $\beta$ to be diagonal, $\beta = diag(\beta_1,\beta_2,...,\beta_m)$, and in numerical simulations only one component $\beta_1$ is assumed to be nonzero. $\gamma$ is a column matrix ($m \times 1$), with elements zero or $1$, and it decides the components of $x_1$ and $x_2$ that gets feedback from the environment. $\gamma^{T}$ is the transpose of $\gamma$ and it decides the components of $x_1$ and $x_2$ that give feedback to the environment. We take $\varepsilon_d$ to be the strength of direct diffusive coupling between the systems, and $\varepsilon_e$  the strength of feedback coupling between the systems and the environment. 

The direct coupling $\varepsilon_d$, gives a synchronizing tendency between the two systems while the coupling through the environment $\varepsilon_e$, gives an anti-synchronizing tendency. Thus, when both the couplings are above their critical values, there is a competition between the two tendencies, and the net result is the amplitude death. In the amplitude death state, the coupled systems are driven to a fixed point.

\begin{figure}
\includegraphics[width=0.95\columnwidth]{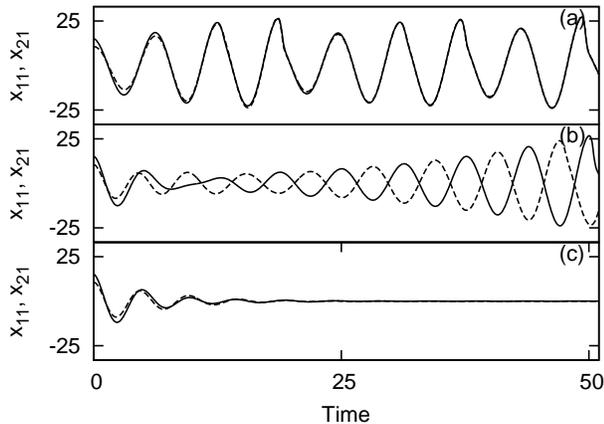}
\caption{ \label{rosts} Time series of the first variables $x_{i1},i=1,2$ of two coupled R\"ossler systems. (a) Synchronization for $(\varepsilon_d,\varepsilon_e) = (0.2,0.0)$.(b) Anti-phase synchronization $(\varepsilon_d,\varepsilon_e) = (0.0,1.0)$.(c) Amplitude death for $(\varepsilon_d,\varepsilon_e) = (0.2,1.0)$. Here, the R\"ossler parameters are $a = b = 0.1, c=18$. The damping parameter of the environment is taken to be $\kappa=1$. }
\end{figure}

We illustrate the above scheme for two coupled chaotic R\"ossler systems represented by the following equations ($i,j=1,2, \; i\ne j$):
\begin{eqnarray}
\dot{x}_{i1} & = & -x_{i2} - x_{i3} + \varepsilon_d (x_{j1} - x_{i1}) + \varepsilon_e y, 
\nonumber \\
\dot{x}_{i2} & = & x_{i1} + a x_{i2}, 
\nonumber \\
\dot{x}_{i3} & = & b + x_{i3} ( x_{i1} -c ), 
\nonumber \\
\dot{y} & = & -\kappa y - \frac{\varepsilon_e}{2} \sum_{i}{ x_{i1}}.
 \label{eq:rossler}
\end{eqnarray}
The resulting time series for a synchronized state with only direct coupling, an anti-phase-synchronized state with only indirect coupling, and the amplitude death state with both direct and indirect couplings  are shown in Fig.~\ref{rosts}. When $\varepsilon_e=0$, and $\varepsilon_d$ is sufficiently large, we observe synchronization [Fig.~\ref{rosts}(a)]. When $\varepsilon_e$ is increased for $\varepsilon_d=0$, the systems are in an anti-phase synchronized state [Fig.~\ref{rosts}(b)]. When both $\varepsilon_e$ and $\varepsilon_d$ are sufficiently large, the systems stabilize to a state of amplitude death [Fig.~\ref{rosts}(c)]. 

\begin{figure}
\includegraphics[width=0.95\columnwidth]{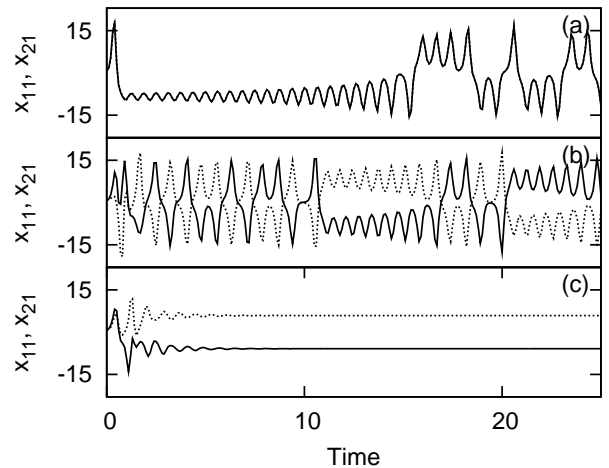}
\caption{ \label{lorts} Time series of the first variables $x_{i1}$ of the coupled Lorenz systems for the parameters $(\sigma = 10, r = 28, b = 8/3)$ and $\kappa=1$. (a) Synchronization for $(\varepsilon_d,\varepsilon_e) = (5,0)$. (b) Anti-synchronization for $(\varepsilon_d,\varepsilon_e) = (0,12)$. (c) Amplitude death for $(\varepsilon_d,\varepsilon_e) = (5,12)$. }
\end{figure}

We apply the same scheme to two coupled chaotic Lorenz systems as given by the following equations: 
\begin{eqnarray}
\dot{x}_{i1} & = & \sigma (x_{i2}-x_{i1}) + \varepsilon_d (x_{j1} - x_{i1}) + \varepsilon_e y, 
\nonumber \\
\dot{x}_{i2} & = & (r-x_{i3})x_{i1} -x_{i2}, 
\nonumber \\
\dot{x}_{i3} & = & x_{i1} x_{i2} - b x_{i3}, 
\nonumber \\
\dot{y} & = & -\kappa y - \frac{\varepsilon_e}{2} \sum_{i}{ x_{i1}}.
 \label{eq:lorenz}
\end{eqnarray}
We find that amplitude death occurs in this case also. This is illustrated in Fig.~\ref{lorts}, where time series for a synchronized state [Fig.~\ref{lorts}(a)], anti-synchronized state [Fig.~\ref{lorts}(b)]  and an amplitude death state [Fig.~\ref{lorts}(c)] are shown. 

\begin{figure}
\includegraphics[width=0.95\columnwidth]{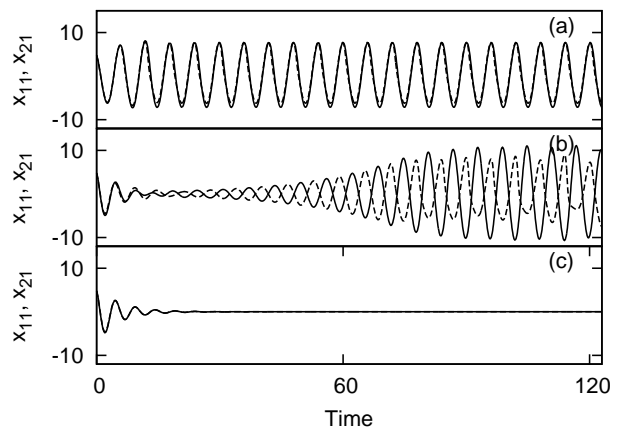}
\caption{ \label{rostspm} Time series of the first variables $x_{i1},i=1,2$ of two coupled non-identical R\"ossler systems. Here, the parameters $a$ and $b$ of the two R\"ossler systems are chosen to be same ($a=b=0.1$), while the parameters $c$ of the systems are chosen such that one of the systems is in a chaotic state ($c=18$) and the other in a periodic state ($c=4$). The damping parameter of the environment is chosen to be $\kappa=1$. (a) Generalized synchronization for $(\varepsilon_d,\varepsilon_e) = (1.0,0.0)$. (b) Anti-phase-synchronization for $(\varepsilon_d,\varepsilon_e) = (0.0,1.0)$. (c) Amplitude death for $(\varepsilon_d,\varepsilon_e) = (1.0,1.0)$.}
\end{figure}

So far we have presented the method for identical systems. However, the method
also works for nonidentical systems. In general, for nonidentical systems 
the direct coupling will give a generalized synchronization between the
coupled systems. Similarly, the anti-synchronization due to the
indirect coupling will also become of a generalized type. The combination of 
direct and indirect coupling still leads to amplitude death. As an example 
consider two coupled R\"ossler systems. In Eq.~(\ref{eq:rossler}), we keep the 
parameters of one system fixed and vary the parameter $c$ of the other 
system. We find that for sufficient strength of coupling the systems go to 
the amplitude death state even for large deviations in $c$. 
Also, the amplitude death state occurs when the 
individual non-interacting systems are in different dynamical regimes. This is shown in Fig.~\ref{rostspm}, where time series for a generalized synchronized state [Fig.~\ref{rostspm}(a)], anti-phase synchronized state [Fig.~\ref{rostspm}(b)]  and an amplitude death state [Fig.~\ref{rostspm}(c)] are shown for two non-identical R\"ossler systems. 
\section{Linear stability analysis}
We present an analysis of the stability of the steady state of two systems coupled via the scheme given in Eq.~(\ref{eq:model}). For this, we write the variational equations  formed by linearizing Eq.~(\ref{eq:model}) as
\begin{eqnarray}
\dot{\xi}_1 & = & f'(x_1) \xi_1 + \varepsilon_d \beta (\xi_2 - \xi_1)+ \varepsilon_e \gamma z, \nonumber \\
\dot{\xi}_2 & = & f'(x_2) \xi_2 + \varepsilon_d \beta (\xi_1 - \xi_2)+ \varepsilon_e \gamma z, \nonumber \\
\dot{z} & = & -\kappa z - \frac{\varepsilon_e}{2} \gamma^{T} (\xi_1+\xi_2),
\label{eq:deviation}
\end{eqnarray}
where $\xi_1$, $\xi_2$ and $z$ are small deviations from the respective values. We denote synchronizing and anti-synchronizing tendencies through the variables $\xi_s$ and $\xi_a$ respectively  as given by 
\begin{eqnarray}
\xi_s & = & \xi_1 - \xi_2, \nonumber \\
\xi_a & = & \xi_1 + \xi_2. 
\label{def:sanda}
\end{eqnarray}
Then Eq.~(\ref{eq:deviation}) can be written as
\begin{eqnarray}
\dot{\xi}_s & = & \frac{f'(x_1)+f'(x_2)}{2} \xi_s + \frac{f'(x_1)-f'(x_2)}{2} \xi_a - 2 \varepsilon_d \beta \xi_s, \nonumber \\
\dot{\xi}_a & = & \frac{f'(x_1)-f'(x_2)}{2} \xi_s + \frac{f'(x_1)+f'(x_2)}{2} \xi_a + 2 \varepsilon_e \gamma z, \nonumber \\
\dot{z} & = & -\kappa z - \frac{\varepsilon_e}{2} \gamma^{T} \xi_a.
\label{eq:saz}
\end{eqnarray}
For stability, all the Lyapunov exponents obtained from Eq.~(\ref{eq:saz}) should be negative.

In general, it is not easy to analyze the stability of the synchronized state from Eq.~(\ref{eq:saz}). However, considerable progress can be made if we assume that the time average values of $f'(x_1)$ and $f'(x_2)$ are approximately the same and can be replaced by an effective constant value $\mu$. In this approximation we treat $\xi_1$ and $\xi_2$ as scalars. This approximation simplifies the problem such that only the relevant features remain and is expected to give features near the transition. This type of approximation was used in Refs.~\cite{amb09,Res10} and it was noted that it describes the overall features of the phase diagram reasonably well. Thus, Eq.~(\ref{eq:deviation}) becomes
\begin{subequations}
\label{eq:sazmu}
\begin{eqnarray}
\label{eq:sazmu1}
\dot{\xi}_s & = &\mu \xi_s - 2 \varepsilon_d \xi_s, \\ 
\label{eq:sazmu2}
\dot{\xi}_a & = & \mu \xi_a + 2 \varepsilon_e z, \\
\label{eq:sazmu3}
\dot{z} & = & -\kappa z - \frac{\varepsilon_e}{2} \xi_a.
\end{eqnarray}
\end{subequations}
We note that Eqs.~(\ref{eq:sazmu2}) and~(\ref{eq:sazmu3}) are coupled while Eq.~(\ref{eq:sazmu1}) is independent of the other two. The synchronizing tendency is given by Eq.~(\ref{eq:sazmu1}) and the corresponding Lyapunov exponent is
\begin{equation}
\lambda_1 = \mu- 2 \varepsilon_d .
\label{eq:lyap1}
\end{equation}
The anti-synchronizing tendency is given by Eqs.~(\ref{eq:sazmu2}) and~(\ref{eq:sazmu3}). The corresponding Jacobian is
\[ J =  \left( \begin{array}{cc}
\mu & 2 \varepsilon_e \\
-\varepsilon_e/2 & -\kappa \end{array} \right)\]
and the eigenvalues are
\begin{equation}
\lambda_{2,3} = \frac{(\mu - \kappa) \pm \sqrt{(\mu - \kappa)^2 - 4 (\varepsilon_e^2 - \mu \kappa)} }{2}.
\label{eq:lyap23}
\end{equation}
As noted in the preceding section, amplitude death is obtained when both synchronizing and anti-synchronizing tendencies are present and the corresponding coupling constants are greater than the critical values required for the respective phenomena. The synchronizing and anti-synchronizing tendencies become effective when the corresponding Lyapunov exponents, i.e., the real parts of the eigenvalues, are negative. From Eq.~(\ref{eq:lyap1}) we obtain the condition 
\begin{equation}
\varepsilon_d > \mu /2,
\label{eq:stability1}
\end{equation}
while from Eq.~(\ref{eq:lyap23}) we get the following conditions: \\
(1) If $( \mu - \kappa)^2 < 4 (\varepsilon_e^2 - \mu \kappa) $, $\lambda_{2,3}$ are complex and the condition of stability is 
\begin{equation}
\kappa > \mu,
\label{eq:stability2}
\end{equation}
(2) If $( \mu - \kappa)^2 > 4 (\varepsilon_e^2 - \mu \kappa) $, $\lambda_{2,3}$ are real and the stability condition becomes 
\begin{equation}
\kappa > \mu  \; \; \textrm{and} \; \; \varepsilon_e^2 > \mu \kappa .
\label{eq:stability3}
\end{equation}
If Eqs.~(\ref{eq:stability1}) and (\ref{eq:stability2}) or~(\ref{eq:stability3}) are simultaneously satisfied, the oscillations can not occur and the systems stabilize to a steady state  of amplitude death. For a given $\kappa$ and $\mu$, the transition to amplitude death occurs at critical coupling strengths $\varepsilon_{dc}$ and $\varepsilon_{ec}$ that are independent of each other. That is, 
\begin{equation}
\varepsilon_{dc} = const ,
\label{eq:curve1}
\end{equation}
and 
\begin{equation}
\varepsilon_{ec} = const .
\label{eq:curve2}
\end{equation}
These general stability criteria are numerically verified for different systems in the following section.

We can also analyze the stability of amplitude death by noting that the amplitude death corresponds to a fixed point of the coupled system. Thus, the condition for the stability of amplitude death is that all Lyapunov exponents of the fixed point are negative. This can be done for different systems numerically and is discussed in the next section.
\section{Numerical Analysis}
We apply our scheme to two chaotic systems, R\"ossler and Lorenz.
\subsection{Coupled R\"ossler systems}
\begin{figure}
\includegraphics[width=0.95\columnwidth]{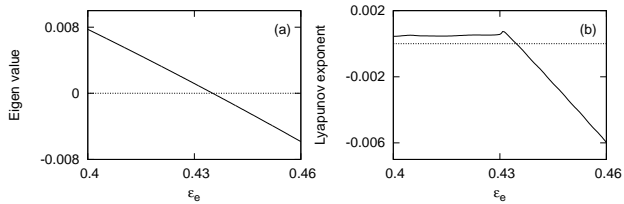}
\caption{ \label{rosle}  (a) Real parts of the largest eigenvalues of the Jacobian obtained from Eq.~(\ref{eq:rossler}), evaluated at the fixed point given in Eq.~(\ref{eq:rosfp}), for increasing $\varepsilon_e$ and fixed values of $\varepsilon_d = 0.2$ and $\kappa = 1$.  At the transition ($\varepsilon_{ec} \sim 0.435$), the real parts of one pair of complex conjugate eigenvalues cross zero. (b) Largest Lyapunov exponent of the coupled R\"ossler system given in Eq.~(\ref{eq:rossler}) for increasing $\varepsilon_e$ for fixed values of $\varepsilon_d = 0.2$ and $\kappa = 1$. The zero crossing of the largest Lyapunov exponent ($\varepsilon_{ec} \sim 0.435$)  indicates the transition to the amplitude death state. In both figures, zero is shown as a dotted line.  }
\end{figure}

Now, we apply the scheme of coupling introduced in Eq.~(\ref{eq:model}) to the case of two chaotic R\"ossler systems. The occurrence of amplitude death in this case is illustrated in Fig.~\ref{rosts}(c). This is further confirmed by calculating the Lyapunov exponents \cite{Wolf85} also.  When the systems are in the amplitude death state, all the Lyapunov exponents of the coupled system are found to be negative. Figure~\ref{rosle}(b) shows the largest Lyapunov exponent of the coupled system as a function of coupling strength $\varepsilon_e$.

\begin{figure}
\includegraphics[width=0.95\columnwidth]{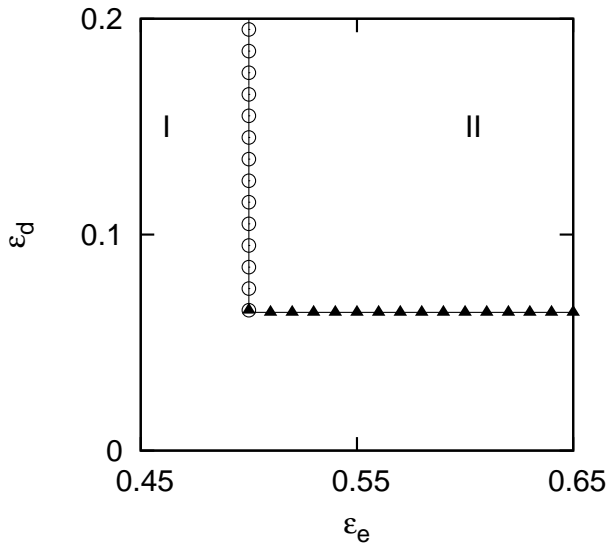} 
\caption{ \label{e1e3ros} Transition from region of oscillations (I) to region of amplitude death (II) in the parameter plane $\varepsilon_e$-$\varepsilon_d$ for coupled R\"ossler systems. Numerical simulations are done with $\kappa = 1$. The points mark the parameter values ($\varepsilon_{ec}$,$\varepsilon_{dc}$) at which the transition to amplitude death occurs. Solid triangles show the transition to amplitude death as $\varepsilon_d$ is increased for a constant $\varepsilon_e$. The horizontal line formed by these triangles confirms the stability condition Eq.~(\ref{eq:curve1}). Similarly, the circles correspond to transition to the amplitude death state as $\varepsilon_e$ is increased for a constant $\varepsilon_d$ and confirm the stability condition of Eq.~(\ref{eq:curve2}). }
\end{figure}

We study the transition to death by identifying regions of amplitude death in the parameter plane of coupling strengths $\varepsilon_e-\varepsilon_d$ for a chosen value of $\kappa$. To characterize the state of amplitude death, we use an index $A$, defined as the difference between the global maximum and global minimum values of the time series of the system over a sufficiently long interval. The case where $A=0$ represents the state of amplitude death, while $A \neq 0$ indicates oscillatory dynamics.  The parameter value at which $A$ becomes $\sim 0$ is thus identified as the threshold for onset of stability of amplitude death states. Using this index, the transition curves in the parameter plane $\varepsilon_e - \varepsilon_d$ are plotted in Fig.~\ref{e1e3ros}. We note that the points obtained from numerical simulations agree with the stability criteria Eqs.~(\ref{eq:curve1}) and (\ref{eq:curve2}) obtained in the preceding section.

\begin{figure}
\includegraphics[width=0.95\columnwidth]{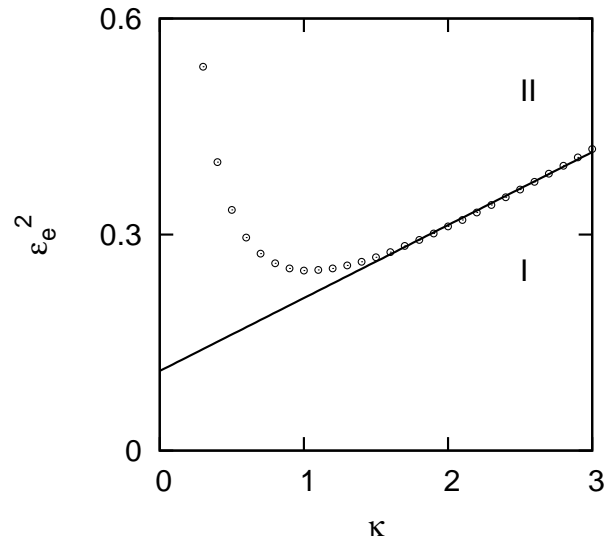}
\caption{ \label{kape1sqros} Transition from region of oscillations (I) to region of amplitude death (II) in the parameter plane $\kappa-\varepsilon_e^{2}$ for the coupled R\"ossler system. The points of amplitude death are obtained numerically when the index $A$ becomes zero. The solid curve is a linear fit corresponding to the stability condition Eq.~(\ref{eq:stability3}), with the effective $\mu = 0.1$. The deviation from straight line behavior for small values of $\kappa$ is discussed in the text.}
\end{figure}

We also verify numerically the criteria for transition to amplitude death given in Eq.~(\ref{eq:stability3}). For this, the numerically obtained values of $\varepsilon_{ec}^2$ are plotted against $\kappa$ in Fig.~\ref{kape1sqros}. The line corresponds to the stability condition Eq.~(\ref{eq:stability3}) and the points are obtained from numerical simulations. It is seen that the agreement is good for larger values of $\kappa$. However, for small values of $\kappa$, the points deviate from straight line behavior.  The reason can be seen from Eq.~(\ref{eq:stability2}) which gives the lower limit on $\kappa$. As $\kappa$ decreases, the damping of the environment variable $y$ is reduced. However, this damping is essential for the anti-synchronizing tendency arising from the coupling to the environment. This leads to the deviations for small values of $\kappa$. 

\begin{figure}
\includegraphics[width=0.95\columnwidth]{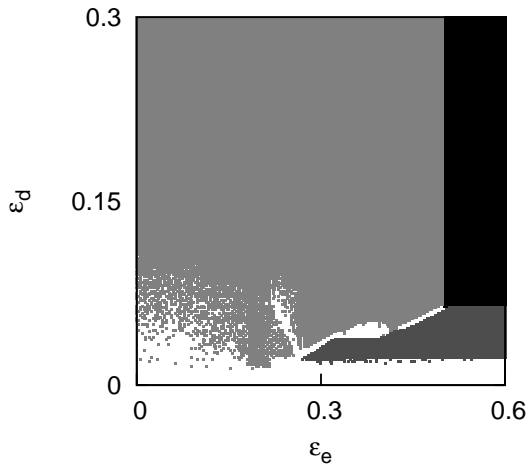}
\caption{ \label{rospp} Regions of different dynamical states in the parameter plane of coupling strengths $\varepsilon_e$--$\varepsilon_d$ in the case of two coupled R\"ossler systems.  The indices used to identify the different regions are the average correlation $C$ [ Eq.~(\ref{eq:corr}) ] and the index $A$. The black region corresponds to the state of amplitude death ($A \sim 0$), the light-gray region to the synchronized state ($C \sim 1$), the dark-gray region to the anti-synchronized state $(C \sim -1)$ and the white region correspond to the state where $|C| < 1$. Here, the parameters are the same as in Fig.~\ref{rosts}.}
\end{figure}

For two coupled R\"ossler systems as given in Eq.~(\ref{eq:rossler}), we study the complete phase diagram in the parameter plane of coupling strengths, identifying the regions of different  dynamic states such as amplitude death, complete  synchronization and anti-synchronization. Amplitude death states are identified using the index $A$ as mentioned above. To identify synchronized or anti-synchronized states we use the asymptotic correlation values as the index, calculated using the equation
\begin{equation}
\label{eq:corr}
C = \frac{ <( x_{11}(t) - <x_{11}(t)> ) ( x_{21}(t) - <x_{21}(t)> )> }{ \sqrt{ <( x_{11}(t) - <x_{11}(t)> )^2> <( x_{21}(t) - <x_{21}(t)> )^2> } }.
\end{equation}
The phase diagram thus obtained for the coupled R\"ossler system is shown in Fig.~\ref{rospp}. When the coupling strengths $\varepsilon_d$ and $\varepsilon_e$ are small, the systems are not synchronized (white region). For small values of $\varepsilon_e$, when $\varepsilon_d$ is increased, the systems synchronize (light-gray region). When $\varepsilon_e$ is increased, the systems become anti-synchronized (dark-gray region). When both the coupling strengths are above a certain threshold as given by the stability conditions Eqs.~(\ref{eq:curve1}) and (\ref{eq:curve2}), the systems stabilize to the state of amplitude death (black region). We also note that the transition from complete synchronization to anti-synchronization corresponds to a phase transition where the average phase difference between the oscillators changes from $0$ to nearly $\pi$. This is similar to the phase-flip bifurcation reported in the context of time-delay coupled systems \cite{Pra08, Kar10}.

\begin{figure}
\includegraphics[width=0.95\columnwidth]{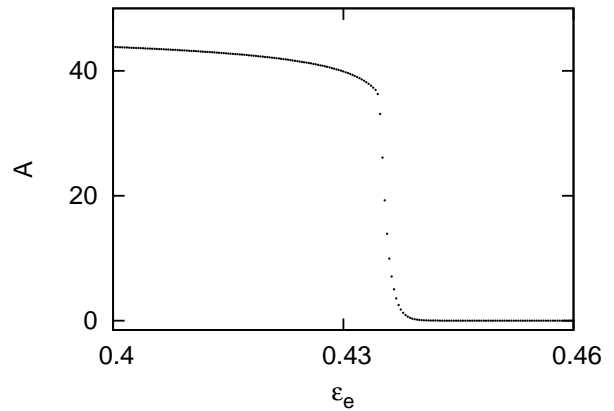}
\caption{ \label{rostransition} The index $A$ as a function of $\varepsilon_e$ for a fixed value of $\varepsilon_d = 0.2$ for two coupled R\"ossler systems. As $\varepsilon_e$ is increased, we observe a continuous transition to the state of amplitude death $(\varepsilon_{ec} \sim 0.45)$.}
\end{figure}

The nature of the transitions to the state of amplitude death is further characterized by fixing one of the parameters $\varepsilon_e$ or $\varepsilon_d$ and increasing the other. This is shown in Fig.~\ref{rostransition}, where the index $A$ is plotted for increasing $\varepsilon_e$ for a chosen value of $\varepsilon_d$. Here, the transition from the oscillatory to the amplitude death state is continuous such that, as the coupling strength is increased, the amplitude of the oscillations gradually decreases to zero. A similar transition is observed for the case where $\varepsilon_e$ is kept fixed and $\varepsilon_d$ is increased.

\begin{figure}
\includegraphics[width=0.95\columnwidth]{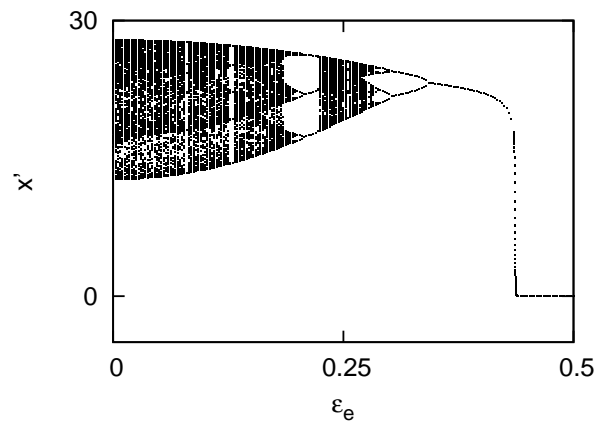}
\caption{ \label{rosbif} Bifurcation diagram  obtained by plotting the maxima of $x_{11}$ ($x'$) for sufficiently long period by increasing the coupling strength $\varepsilon_e$ for $\varepsilon_d = 0.2$ and $\kappa=1$. Here, the systems go through a reverse period-doubling bifurcation as $\varepsilon_e$ increases, leading to a one-cycle (at $\varepsilon_{e} \sim 0.35$)  before amplitude death occurs $(\varepsilon_{ec} \sim 0.45)$.}
\end{figure}

We also notice from the time series and phase space plot that, as the coupling strength increases ($\varepsilon_d$ or $\varepsilon_e$), the R\"ossler systems undergo the full reverse period-doubling sequence to the one-cycle state before going to the amplitude death state. Then, the transition to the state of amplitude death occurs via a supercritical Hopf bifurcation. The bifurcation diagram for this transition is shown in Fig.~\ref{rosbif}.  
This is further confirmed by computation of the fixed points of the coupled system and their stability near the transition region. Numerical simulations show that the coupled R\"ossler systems in Eq.~(\ref{eq:rossler}) stabilize to the steady state corresponding to synchronized states of the subsystems. These synchronized steady states are obtained from Eq.~(\ref{eq:rossler}) as
\begin{eqnarray}
\label{eq:rosfp}
x_{i1}^{*} & = & ( c \pm \sqrt{c^2-  4ab\kappa / (\kappa-\varepsilon_e^{2}a) })/2, \nonumber \\
x_{i2}^{*} & = & -x_{i1}^{*}/a , \nonumber \\
x_{i3}^{*} & = & -b/(x_{i1}^{*}-c) , \nonumber \\
y^{*} & = & -\varepsilon_e x_{i1}^{*}/ \kappa . 
\end{eqnarray}

Of the two fixed points, the one with the plus sign in the second term of the $x_{i1}^*$ equation is unstable and the one with the minus sign in the second term of $x_{i1}^*$ equation become stable in the amplitude death state. The nature of the transition to the stable fixed point is determined by the eigenvalues of the corresponding Jacobian, and we find that at the transition, real parts of the complex conjugate pairs of eigenvalues become negative [Fig.~\ref{rosle}(a)] , indicating a supercritical Hopf bifurcation as described in Ref.\cite{Strogatz1994}.  In the amplitude death region, all the Lyapunov exponents of the coupled system [ given in Eq.~\ref{eq:rossler}] are  found to be negative. The largest Lyapunov exponent of the coupled system crosses zero at the transition, and this is shown in Fig.~\ref{rosle}(b).  The nature of the transition is found to be the same when $\varepsilon_e$ is kept fixed and $\varepsilon_d$ is increased.

The above numerical results are presented for one set of parameters of the
R\"ossler system. We have varied the parameters and verified that the method
works for other values of the parameters.
\subsection{Coupled Lorenz systems}
\begin{figure}
\includegraphics[width=0.95\columnwidth]{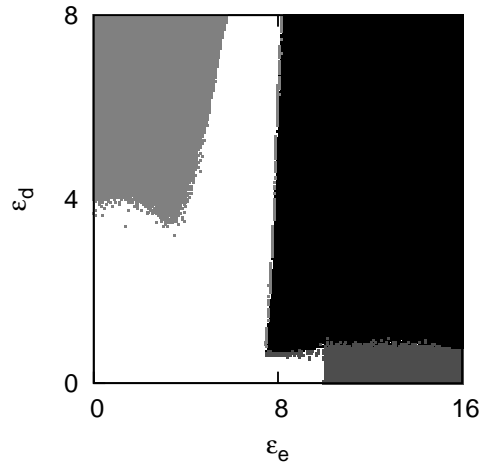}
\caption{ \label{lorpp} Regions of different dynamical states in the parameter plane of coupling strengths $\varepsilon_e$--$\varepsilon_d$ in the case of Lorenz systems. The black region corresponds to the state of amplitude death ($A \sim 0$), the light-gray region corresponds to the synchronized state ($C \sim 1$), the dark-gray region correspond to the anti-synchronized state $(C \sim -1)$ and the white region correspond to the state where $|C| < 1$.}
\end{figure}

\begin{figure}
\includegraphics[width=0.95\columnwidth]{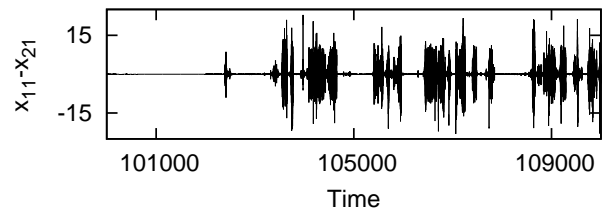}
\caption{ \label{errlor} Synchronization error ($x_{11}-x_{21}$) as a function of time in the case of Lorenz systems for $\varepsilon_d=5$ and $\varepsilon_e=4.665$.}
\end{figure}

\begin{figure}
\includegraphics[width=0.95\columnwidth]{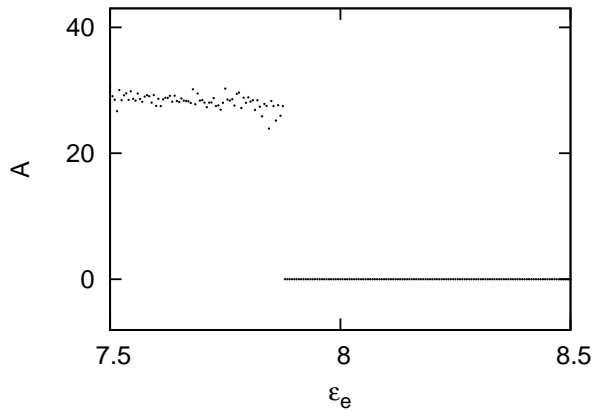}
\caption{ \label{lortransition} The index $A$ as a function of $\varepsilon_e$ for a fixed value of $\varepsilon_d = 5.0$ for two coupled Lorenz systems. Note that the transition to amplitude death is sudden ($\varepsilon_{ec} \sim 7.88$) as opposed the case of coupled R\"ossler systems where the transition is continuous.}
\end{figure}

We repeat the same study in the case of two coupled Lorenz systems. It is interesting to note that, in this case, the coupled systems stabilize to a fixed point that corresponds to anti-synchronized states for the subsystems ($x_{11}=-x_{21},x_{12}=-x_{22},x_{13}=x_{23}$) as shown earlier in Fig.~\ref{lorts}(c). The  regions of different dynamical states in the parameter plane of coupling strengths in this case are shown in Fig.~\ref{lorpp}. 
When both $\varepsilon_d$ and $\varepsilon_e$ are small, the systems are not synchronized (white region). For very small values of $\varepsilon_d$ and large $\varepsilon_e$, the systems are anti-synchronized (dark-gray), and when  $\varepsilon_d$ is increased from this state, the systems go to the amplitude death state (black region). For small values of $\varepsilon_e$ and large $\varepsilon_d$, the systems are synchronized (light-gray). As $\varepsilon_e$ increases, the systems first lose synchronization, and for larger values of $\varepsilon_e$, they stabilize to the state of amplitude death (black). In the de-synchronized state before the amplitude death state, the attractor in the phase space is highly distorted and the system goes through a state of frustration, trying to stabilize to the anti-synchronized state from the synchronized state before death occurs. This is illustrated in Fig.~\ref{errlor}, where the time series of the synchronization error between the two Lorenz systems is shown. Near this transition region, some initial conditions remain in a chaotic transient state for a long time before becoming stabilized to the fixed point. The phenomena of  multi-stability and hysteresis are also observed in this region. 
 The nature of the transition to amplitude death in this case is shown in Fig.~\ref{lortransition}. Unlike the case of coupled  R\"ossler systems, here we see that the amplitude of oscillations drops suddenly at a critical strength of coupling.  Thus, the transition is directly from the chaotic to the amplitude death state. A similar type of transition from the chaotic to the amplitude death state in the case of time-delay coupled Lorenz systems has been reported in Ref.~\cite{Pra05}. We further characterize this transition by computing the fixed points of the coupled system given in Eq.~(\ref{eq:lorenz}) and evaluating their stability near the transition region.

\begin{figure}
\includegraphics[width=0.95\columnwidth]{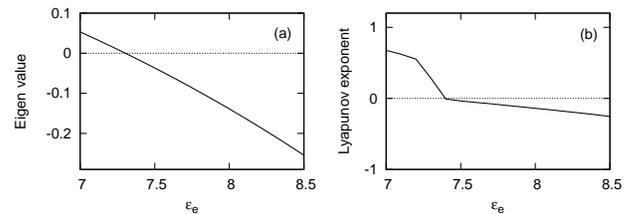}
\caption{ \label{lorle} (a) Real parts of the largest eigenvalues of the Jacobian obtained from Eq.~(\ref{eq:lorenz}), evaluated at the fixed point given in Eq.~(\ref{eq:lorfp}), for increasing $\varepsilon_e$ and fixed values of $\varepsilon_d = 5$ and $\kappa = 1$.  At the transition ($\varepsilon_{ec} \sim 7.31$), the real parts of one pair of complex conjugate eigenvalues cross zero. (b) Largest Lyapunov exponent of the coupled Lorenz system given in Eq.~(\ref{eq:lorenz})for increasing $\varepsilon_e$ for fixed values of $\varepsilon_d = 5$ and $\kappa = 1$. The zero crossing of the largest Lyapunov exponent ($\varepsilon_{ec} \sim 7.43$)  indicates transition to the amplitude death state. In both figures, zero is shown as a dotted line. }
\end{figure}

Numerical simulations show that the coupled Lorenz systems in Eq.~(\ref{eq:lorenz}) stabilize to the steady state corresponding to the anti-synchronized state of the subsystems. These steady states are obtained from Eq.~(\ref{eq:lorenz}) as
\begin{eqnarray}
\label{eq:lorfp}
x_{11}^{*} & = & \pm \sqrt{\frac{((r-1) \sigma-2 \varepsilon_d) b}{\sigma + 2 \varepsilon_d}}, \nonumber \\
x_{12}^{*} & = & \frac{(\sigma+ 2 \varepsilon_d)}{ \sigma} x_{11}^{*},  \nonumber \\
x_{13}^{*} & = & \frac{(\sigma+ 2 \varepsilon_d)}{ \sigma b} x_{11}^{*2},  \nonumber \\
x_{21}^{*} & = & - x_{11}^{*}, \nonumber \\
x_{22}^{*} & = & - x_{12}^{*}, \nonumber \\
x_{23}^{*} & = & x_{13}^{*}, \nonumber \\
y^{*} & = & 0 .
\end{eqnarray} 
For both the solutions, we find that at the transition, the real parts of the complex conjugate pairs of eigenvalues of the corresponding Jacobian become negative [Fig.~\ref{lorle}(a)]. For $\varepsilon_d > \varepsilon_{dc}$, numerically an unstable limit cycle is found to coexist with the stable state of amplitude death for certain initial values. As there is no stable limit cycle before amplitude death, and an unstable fixed point becomes stable, it seems that this is a sub-critical Hopf bifurcation. All the Lyapunov exponents of the system [Eq.~\ref{eq:lorenz}] are found to be negative at the amplitude death state. At the transition, the largest Lyapunov exponent of the coupled system becomes negative as shown in Fig.~\ref{lorle}(b). 
\section{Amplitude death in other cases}
We have presented a mechanism for inducing amplitude death in coupled systems due to the competing effects of synchronizing and anti-synchronizing tendencies. We have shown this in the context of two chaotic systems, namely, coupled R\"ossler and coupled Lorenz systems. To test the generality of the method, we apply it to a variety of systems and find that, for all cases which are synchronizable, the method works. We understand that extensive numerical simulations may not analytically establish the generality of the mechanism. However, the results of our numerical simulations appears to indicate that this method is quite general. 

We also give an intuitive physical argument to support our claim of generality of the method. Our mechanism consists of having two types of coupling. The first is direct coupling, which leads to synchronization. We know that, if the coupling constant is sufficiently large, the synchronization condition ensures that the largest Lyapunov exponent transverse to the synchronization manifold is negative. Considering the space of coupled oscillators as a product of the individual systems and a network of two nodes, the synchronization manifold corresponds to the direction $e_s = (1,1)^T$ in the network coordinates. Similarly, the coupling through the environment which ensures anti-synchronization leads to the condition that the largest Lyapunov exponent transverse to the direction $e_a = (1, -1)^T$ is negative. Since we have coupled only two systems, ensuring that the largest exponents transverse to both synchronizing and anti-synchronizing directions, i.e. $e_s$ and $e_a$, are negative, implies that all the Lyapunov exponents are negative. Thus the system must converge to a fixed point.  

 In this section, we present the results of applying this method to periodic, time-delay, hyperchaotic, and driven systems and different schemes of direct coupling.
\subsection{Amplitude death in periodic systems}\label{ap:periodic}
We study two standard limit cycle oscillators, namely Landau-Stuart and van der Pol oscillators, coupled using the scheme given in Eq.~(\ref{eq:model}). The Landau-Stuart system is a nonlinear limit cycle oscillator, which has been  previously used as a model system for studying the phenomenon of amplitude death \cite{Kar07,Ram98}.  In our case, the dynamics of two coupled Landau-Stuart systems is given by the following set of equations:
\begin{eqnarray}
\dot{x}_{i1} & = & (x_{i1}^2 + x_{i2}^2)x_{i1} - \omega x_{i2} + \varepsilon_d  (x_{j1} - x_{i1}) + \varepsilon_e y, 
\nonumber \\
\dot{x}_{i2} & = & (x_{i1}^2 + x_{i2}^2)x_{i2} + \omega x_{i1},
\nonumber \\
\dot{y} & = & -\kappa y - \frac{\varepsilon_e}{2} \sum_{i}{ x_{i1}}.
 \label{eq:landau}
\end{eqnarray}
From  numerical analysis of the above equations with $\omega=2$, we see that, for small values of $\varepsilon_e$ and $\varepsilon_d$, the systems are synchronized. For small values of $\varepsilon_d$ and large values of $\varepsilon_e$, the systems are in the anti-synchronized state. When the strengths of both $\varepsilon_d$ and $\varepsilon_e$ are sufficiently large, we observe amplitude death. We note that here the transition to the state of amplitude death is sudden, similar to the case of Lorenz systems. However, in this case, for a given strength of coupling, the stability of amplitude death state depends on the initial conditions indicating multi-stability, that is, some initial conditions go to the amplitude death state, while some other initial conditions remain in the oscillatory state.  Such a multi-stability has also been reported for amplitude death phenomena in the case of Landau-Stuart oscillators using conjugate coupling \cite{Kar07}. For some initial values,  amplitude death occurs even in the absence of direct coupling ( $\varepsilon_d=0$) . A possible explanation is that the $\mu$ of the individual system is negative or zero such that, the stability condition $\kappa > \mu$ given in Eq.~(\ref{eq:stability2}) is always satisfied. It is seen that the area of the basin of amplitude death increases as $\varepsilon_d$ is increased. 

We repeat the same study for the case of two coupled periodic van der Pol systems given by the following equations:
\begin{eqnarray}
\dot{x}_{i1} & = & x_{i2} + \varepsilon_d (x_{j1} - x_{i1}) + \varepsilon_e y, 
\nonumber \\
 \dot{x}_{i2} & = & \alpha (1-x_{i1}^2) x_{i2} - x_{i1},
\nonumber \\
\dot{y} & = & -\kappa y - \frac{\varepsilon_e}{2} \sum_{i}{ x_{i1}}.
\label{eq:vanderPol}
\end{eqnarray}

We choose the parameter $\alpha=1$ such that the system has a stable limit cycle when both the couplings are absent (i.e., $\varepsilon_e=\varepsilon_d =0$). For a suitable strength of the direct coupling $\varepsilon_d$, the systems synchronize and amplitude death is observed when both direct  and indirect couplings are above their respective thresholds. As far as the nature of the transition to the state of amplitude death is concerned, the van der Pol oscillator shows a different behavior from that of other systems. We find that the nature of the transition to amplitude death depends on the type of coupling parameter. Fixing $\varepsilon_d$ and increasing $\varepsilon_e$, we see a smooth transition similar to that in the case of R\"ossler systems  and by fixing $\varepsilon_e$ and increasing $\varepsilon_d$, we get a sudden transition, as in the case of Lorenz systems. 
\subsection{Amplitude death in time delay systems}\label{ap:mg}
 The Mackey-Glass time-delay system is well studied as a model exhibiting hyperchaos. Stabilization to fixed point, or amplitude death in such systems has been reported by use of stabilization methods such as conventional feedback, tracking filters and delayed feedback \cite{Nam95}. Here we consider two Mackey-Glass systems coupled via both direct and indirect couplings as given by
\begin{eqnarray}
\dot{x}_{i} &=& -\alpha x_{i} + \frac{\beta x_{\tau i}}{1+x_{\tau i}^{m}} + {\varepsilon_d} (x_{j} - x_{i}) + \varepsilon_e y, \nonumber \\ 
\dot{y} &=& -\kappa y - \frac{\varepsilon_e}{2} \sum_{j = 1,2} x_j,
\label{eq:mac}
\end{eqnarray}
where $x_i$ represents the Mackey-Glass system \cite{Mac77} and $x_{\tau i}$ is the  value of the variable $x_i$ at a delayed time $x_i(t-\tau)$.  The parameters of the Mackey-Glass systems are chosen such that, the individual systems are in the hyperchaotic regime ( $\alpha = 1$, $\beta = 2$, $\tau = 2.5$, $m=10$). For very small values of $\varepsilon_d$ and $\varepsilon_e$, the two systems are not synchronized. The systems synchronize as $\varepsilon_d$ is increased. For large values of $\varepsilon_d$, if $\varepsilon_e$ is increased, the systems go to a state of amplitude death. When $\varepsilon_d$ is small and $\varepsilon_e$ is large, the systems are in an anti-synchronized state. As we increase $\varepsilon_d$, the systems go to the amplitude death state. However, for large values of $\varepsilon_d$ and $\varepsilon_e$, the Mackey-Glass systems appears to show a different behavior from the R\"ossler or Lorenz systems. We observe a re-entrant behavior to rhythmogenesis, both as $\varepsilon_e$ increases and as $\varepsilon_d$ increases. This transition also satisfies our stability conditions Eqs.~(\ref{eq:curve1}) and (\ref{eq:curve2}). We find that the transition to the state of amplitude death in the case of two coupled Mackey-Glass systems is continuous and that the systems go through a reverse period-doubling sequence reaching a limit cycle before the amplitude death occurs. This is similar to the case of R\"ossler systems discussed earlier.  
\subsection{Amplitude death with alternate schemes for direct coupling}\label{ap:alt_coupling}
So far, we have studied direct coupling of the diffusive type. Synchronization is also possible with direct coupling of different types. Here, we now study two such types of coupling.
\subsubsection{Lorenz systems with replacement coupling}\label{ap:lorrep}
Here we consider Lorenz systems coupled using a different scheme of coupling, namely replacement coupling, as given by  the following equations
\begin{eqnarray}
\dot{x}_{i1} & = & \sigma (x_{j2}-x_{i1}) +  \varepsilon_e y, 
\nonumber \\
\dot{x}_{i2} & = & (r-x_{i3})x_{i1} -x_{i2} ,
\nonumber \\
\dot{x}_{i3} & = & x_{i1} x_{i2} - b x_{i3}, 
\nonumber \\
\dot{y} & = & -\kappa y - \frac{\varepsilon_e}{2} \sum_{i}{ x_{i1}}.
 \label{eq:lor_rep}
\end{eqnarray}

Here, the direct coupling is of the replacement type, such that the $x_{2}$ variable in the first function of the first system is that of the second system, and vice versa. This type of coupling leads to synchronization as reported in Ref.~\cite{Gue95}. We introduce indirect coupling through the variable $y$. We find that, for suitable values of the coupling strength $\varepsilon_e$, the systems stabilize to a state of amplitude death.
\subsubsection{Synaptically coupled Hindmarsh-Rose model of neurons}\label{ap:hr}
The Hindmarsh-Rose system is a model of neurons which shows the spiking and bursting behavior of the membrane potential of a single neuron \cite{Hin84}. We take two neurons with excitatory synaptic coupling \cite{Bel05} between them and introduce an indirect coupling as given by the following equations:
\begin{eqnarray}
\dot{x}_{i1}& =& x_{i2} - x_{i1}^{3} + a x_{i1}^2 - x_{i3} + I + \nonumber \\
& & \varepsilon_e y + \varepsilon_d \frac{V_r-x_{i1}}{1+\exp(-\lambda(x_{j1}-\theta))},  \nonumber \\
\dot{x}_{i2}& =& 1-b x_{i1}^2 - x_{i2}, \nonumber \\
\dot{x}_{i3} &  = & \rho( s (x_{i1} + \chi) - x_{i3}), \nonumber \\
\dot{y} &  = &  -\kappa y - \frac{\varepsilon_e}{2} \sum_i {x_{i1}}.
\label{eq:hr}
\end{eqnarray}

\begin{figure}
\includegraphics[width=0.95\columnwidth]{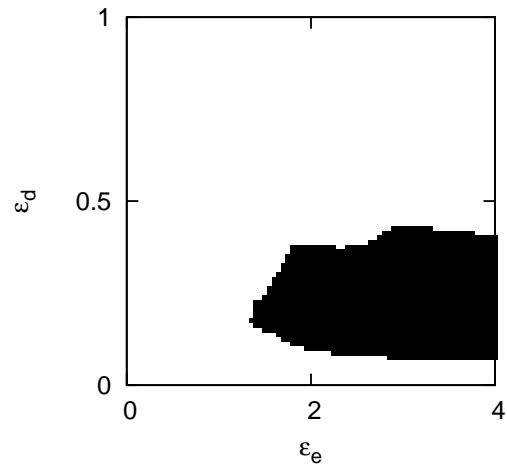}
\caption{ \label{hrpp} Region of amplitude death (black) in the parameter plane of coupling strengths $\varepsilon_e$--$\varepsilon_d$ for two coupled Hindmarsh-Rose neurons. We choose the parameters of the system to be $a=3$,$b=5$, $\rho=0.006$, $s=4$, $\chi=1.6$, and $I=3.2$. The parameters in the coupling terms are chosen to be $ V_r = 3$, $\theta = -0.25$, $\lambda=10$, and $\kappa = 1$. For the neuronal systems, the criterion used in the text for identifying the amplitude death state, i.e., the index $A \sim 0$, needs to be changed. Here, we use the maximum voltage from both neurons ($x_m$) as the index to identify the death state. The criteria for synchronized and anti-synchronized states also needs modification. Hence, the other regions are not shown explicitly in the figure.}
\end{figure}

Here, the variable $x_{i1}$ represents the membrane potential of a neuron and the variables $x_{i2}$ and $x_{i3}$ are related to ion currents across the membrane. We choose the parameters of the system such that the individual neurons are in the chaotic bursting state. When $\varepsilon_e=0$, and $\varepsilon_d$ is sufficiently large, the bursts of both neurons become synchronized. For larger values of $\varepsilon_d$ and $\varepsilon_e$, we observe patches of amplitude death (shown in Fig.~\ref{hrpp}). Thus as we keep one of the coupling parameters fixed ($\varepsilon_e$ or $\varepsilon_d$), and increase the other, we observe a transition to the amplitude death state and again re-entrant behavior to spikes. 
\subsection{Amplitude death with alternate dynamics for the environment}
In the previous sections, we have taken the intrinsic dynamics of the environment to be that of an overdamped harmonic oscillator. Here, we show that amplitude death is possible with other intrinsic dynamics for the environment also.

For this, we consider the case of two R\"ossler systems coupled with a dynamic environment, where the intrinsic dynamics of the environment is that of a damped harmonic oscillator. The dynamics of the coupled system in this case is given by 
\begin{eqnarray}
\label{eq:ros_damped_env}
\dot{x}_{i1} & = & -x_{i2} - x_{i3} + \varepsilon_d (x_{j1} - x_{i1}) + \varepsilon_e y_1,
\nonumber \\
\dot{x}_{i2} & = & x_{i1} + a x_{i2}, 
\nonumber \\
\dot{x}_{i3} & = & b + x_{i3} ( x_{i1} -c ), 
\nonumber \\
\dot{y}_1 & = & y_2 - \frac{\varepsilon_e}{2} \sum_{i}{ x_{i1}}, \nonumber \\
\dot{y}_2 & = & - y_2 - \kappa y_1. 
\end{eqnarray}
Here, the variables $y_1$ and $y_2$ represent a two-dimensional environment together forming an underdamped harmonic oscillator. For very weak coupling ($\varepsilon_d \sim 0$,$\varepsilon_e \sim 0$), the two R\"ossler systems are not synchronized. When the coupling strength $\varepsilon_d$ is increased while $\varepsilon_e$ is kept fixed at zero, the systems become synchronized. On the other hand, when $\varepsilon_ e$ is increased while $\varepsilon_d$ is kept fixed at zero, the systems become anti-phase-synchronized. When both  $\varepsilon_e$ and $\varepsilon_d$ are above their respective thresholds, amplitude death is observed. 

We repeat the same study by taking the intrinsic dynamics of the environment as that of an overdamped Duffing oscillator. The equations in this case are
\begin{eqnarray}
\label{eq:ros_duffing_env}
\dot{x}_{i1} & = & -x_{i2} - x_{i3} + \varepsilon_d (x_{j1} - x_{i1}) + \varepsilon_e y,
\nonumber \\
\dot{x}_{i2} & = & x_{i1} + a x_{i2}, 
\nonumber \\
\dot{x}_{i3} & = & b + x_{i3} ( x_{i1} -c ), 
\nonumber \\
\dot{y} & = & y - \kappa y^3 - \frac{\varepsilon_e}{2} \sum_{i}{ x_{i1}}.
\end{eqnarray}
For small values of $\varepsilon_e$ and $\varepsilon_d$, we see that the systems are not synchronized. They become synchronized as $\varepsilon_d$ is increased from this state. For small values of $\varepsilon_d$ and large values of $\varepsilon_e$, the systems are in an anti-phase-synchronized state. When the strengths of both $\varepsilon_d$ and $\varepsilon_e$ are sufficiently large, we observe amplitude death. The phase diagram in this case is qualitatively similar to that given in Fig.~\ref{rospp}.
\subsection{Amplitude death in hyperchaotic R\"ossler systems} 
We also consider the case of two hyperchaotic R\"ossler systems as given by the following equations
\begin{eqnarray}
\label{eq:rosh}
\dot{x}_{i1} &=& -x_{i2} - x_{i3} + \varepsilon_e y_1, \nonumber \\
& & - \varepsilon_d \cos{\theta} (\sin{\theta} (x_{i1} - x_{j1}) + \cos{\theta} (x_{i3} - x_{j3})),  \nonumber \\
\dot{x}_{i2} &=& x_{i1} + a x_{i2} + x_{i4} + \varepsilon_e y_2, \nonumber \\
\dot{x}_{i3} &=& b + x_{i3} x_{i1} \nonumber\\
& & -\varepsilon_d \sin{\theta} (\sin{\theta} (x_{i1} - x_{j1}) + \cos{\theta} (x_{i3} - x_{j3})), \nonumber \\
\dot{x}_{i4} &=& -c x_{i3} + \sigma x_{i4} + \varepsilon_e y_3, \nonumber \\
\dot{y}_1 &=& -\kappa y_1 - \frac{\varepsilon_e}{2} \sum_{i}{ x_{i1}}, \nonumber \\
\dot{y}_2 &=& -\kappa y_2 - \frac{\varepsilon_e}{2} \sum_{i}{ x_{i2}}, \nonumber \\
\dot{y}_3 &=& -\kappa y_3 - \frac{\varepsilon_e}{2} \sum_{i}{ x_{i4}}, 
\end{eqnarray}
where $i,j=1,2$ and $j \neq i$.

\begin{figure}
\includegraphics[width=0.95\columnwidth]{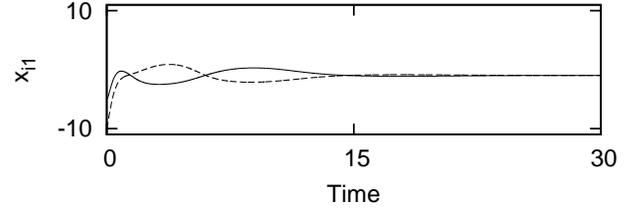}
\caption{ \label{rosh} Time series of the first variables $x_{i1},i=1,2$ of the two coupled hyperchaotic R\"ossler systems given in Eq.~(\ref{eq:rosh}) in the amplitude death state. Parameters of the systems are chosen to be $a=0.25$, $b=3$, $c=0.5$, and $\sigma=0.05$. The parameters used in the direct coupling term are $\varepsilon_d=2.5$, and $\theta=\pi/3$ and in the indirect coupling terms are $\kappa= 10$ and $\varepsilon_e=4$. }
\end{figure}

We choose the parameters of the system such that, the intrinsic dynamics of the systems is hyperchaotic. For this system, the method of time-delay coupling is found ineffective for producing amplitude death\cite{Kon05}. In the absence of coupling via the environment, the direct coupling via a scalar signal results in synchronization of the two hyperchaotic systems for suitable values of parameters $\varepsilon_d$ and $\theta$, as reported in Ref. \cite{Peng96}. We take the environment to be three-dimensional in this case. We find that amplitude death occurs in the coupled system for suitable values of the coupling strengths  $\varepsilon_e$ and $\kappa$. This is shown in Fig.~\ref{rosh}.
\subsection{Small oscillations in driven systems}\label{ap:driven}
We apply the scheme described in this paper to driven systems such as driven van der Pol and Duffing systems. In such driven systems, the fixed point is not a solution for the individual or coupled systems.  So the interpretation of amplitude death as in other systems needs to be changed. Here, we interpret the amplitude death state as the state of very small amplitude oscillations. 

Driven van der Pol systems with direct diffusive coupling and indirect coupling through the environment can be written as
\begin{eqnarray}
\label{eq:dvn}
\dot{x}_{i1} &=& x_{i2} + \varepsilon_d (x_{j1} - x_{i1}) + \varepsilon_e y, \nonumber \\
\dot{x}_{i2} &=& \alpha (1-x_{i1}^2)x_{i2}-x_{i1} + \beta \cos(\omega t), \nonumber \\
\dot{y} &=& -\kappa y - \frac{\varepsilon_e}{2} \sum_j x_{j1}.
\end{eqnarray}

\begin{figure}
\includegraphics[width=0.95\columnwidth]{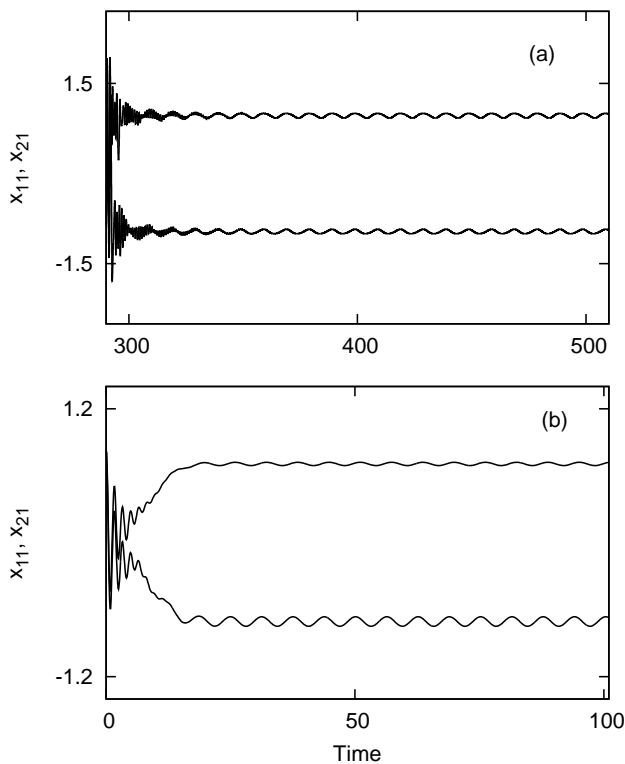}
\caption{ \label{dvnts} Time series of the first variable $x_{i1}$ of the coupled driven systems showing small oscillations. Here, we interpret the small oscillations as the state of amplitude death (see text). (a) Driven van der Pol system for $(\varepsilon_d,\varepsilon_e) = (1.0,3.5)$. The parameters of the individual systems are taken to be $\alpha = 8.53$, $\omega = 0.63$, and $\beta = 1.2$ (b) Driven Duffing system for $(\varepsilon_d,\varepsilon_e) = (1.0,4.0)$. The parameters of the individual systems are taken to be $\alpha = 0.25$, $\omega = 1$, and $\beta = 0.3$}
\end{figure}

We find that, when $\varepsilon_e=0$ and $\varepsilon_d$ increases, the two systems become synchronized. In this state, if we start increasing $\varepsilon_e$, we get a state of small oscillations or amplitude death. Figure ~\ref{dvnts}(a) shows the time series for such a state. 

\begin{figure}
\includegraphics[width=0.95\columnwidth]{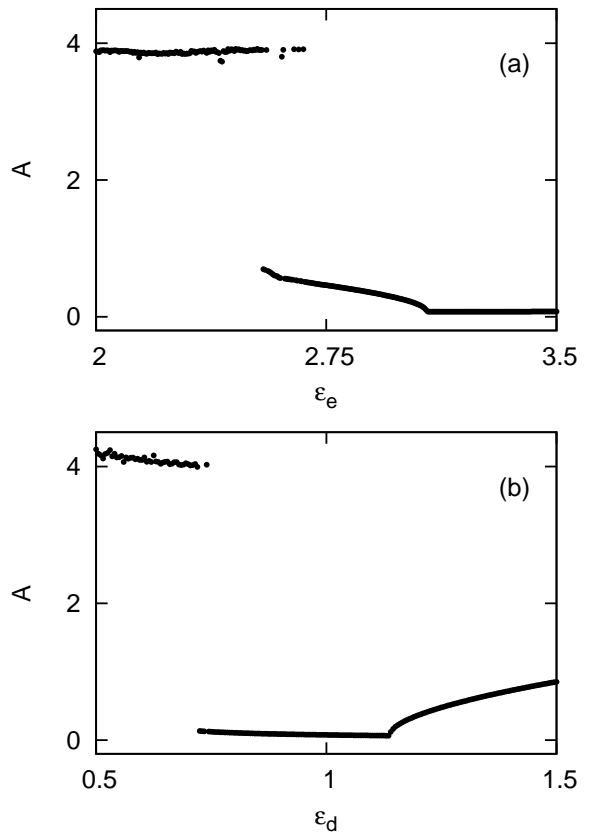}
\caption{ \label{dvntr} Transition to amplitude death state in two coupled driven van der Pol oscillators. Here the death state corresponds to a state of small oscillations since the fixed point is not a stable state. The index $A$ remains finite although very small in the amplitude death state. (a) The index $A$ as a function of $\varepsilon_e$ for fixed $\varepsilon_d=1.0$. The transition is continuous ($\varepsilon_{ec} \sim 3.08$). There is a sudden transition due to a pitchfork bifurcation of the limit cycle observed at $\varepsilon_e \sim 2.55$. (b) The index $A$ as a function of $\varepsilon_d$ for fixed $\varepsilon_e=3.5$. The transition to the amplitude death state is sudden ($\varepsilon_{dc} \sim 0.72$). We also observe a re-entrant continuous transition to periodic oscillations with increasing amplitude at $\varepsilon_d \sim 1.14$.  }
\end{figure}

Figure ~\ref{dvntr}(a) plots the index $A$ as a function of $\varepsilon_e$. We first see a transition from a limit cycle to two different limit cycles for the two systems. This state subsequently goes to the amplitude death state continuously as $\varepsilon_e$ increases further. On the other hand, if we keep $\varepsilon_e$ fixed and increase $\varepsilon_d$, we find a sudden transition to the amplitude death state. As $\varepsilon_d$ increases further, we find a continuous transition to a state of increasing amplitude oscillations. This is shown in Fig.~\ref{dvntr}(b).

The same study is repeated for the case of two coupled Duffing systems given by the following equations:
\begin{eqnarray}
\label{eq:duf}
\dot{x}_{i1} &=& x_{i2} + \varepsilon_d (x_{j1} - x_{i1}) + \varepsilon_e y, \nonumber \\
\dot{x}_{i2} &=& -\alpha x_{i2} + x_{i1} - x_{i1}^3 + \beta \cos(\omega t), \nonumber \\
\dot{y} &=& -\kappa y - \frac{\varepsilon_e}{2} \sum_j x_{j1}.
\end{eqnarray}
We find that a regime of small oscillations is possible  in this case also. This is shown in Fig.~\ref{dvnts}(b).

Thus, we have illustrated that the method for inducing amplitude death in coupled systems introduced in this paper works for periodic, time-delay, hyperchaotic, and driven systems. It is effective in quenching the dynamics even with different forms of direct coupling, such as replacement coupling and synaptic coupling of neurons. 

The context of coupled neurons presents an important case of amplitude death which could explain the mechanism of disruption or suppression of synaptic signals in the case of neuronal disorders, such as Alzheimer's disease, as being due to induced activity and feedback through a protein called amyloid beta (A$\beta$). Using numerical studies on a few neuronal models, we have shown that the competing effects of synaptic activity and the indirect interaction mediated by the protein A$\beta$ lead to sub-threshold activity and synaptic silencing \cite{Res10arXiv}. \section{Discussion}
In this paper, we show that indirect coupling through a dynamic environment in addition to direct coupling can lead to amplitude death in chaotic systems such as R\"ossler and Lorenz systems. The approximate stability analysis developed for general cases gives the transition region in parameter space which is further supported by direct numerical simulations. The nature of the transition to amplitude death is found to be of two typical types, one continuous and the other discontinuous. 

In conclusion, the method  for determining amplitude death introduced in this paper is quite general and works for different types of systems such as periodic, chaotic, hyperchaotic and time-delay systems and also with different types of direct coupling, for example, diffusive, replacement coupling, coupling via scalar signals and synaptic coupling. We have demonstrated that our method works in hyperchaotic R\"ossler systems where time-delay coupling is ineffective in inducing amplitude death. We have also presented a physical argument for the generality of the method. Although the theory is developed for the case of identical systems, we have verified that this method of inducing amplitude death works in the case of non-identical systems as well. In fact, we did not find any exception to our scheme provided the coupled systems are synchronizable.

We also note that the method introduced in this paper can be implemented in practical cases. What is needed is the design of a suitable environment which can introduce the appropriate indirect coupling between the systems.  Moreover, in many natural systems, the environment or external medium exists and can be instrumental in causing suppression of the dynamics. The mechanism presented here provides an explanation for this phenomenon in such cases.
\begin{acknowledgments}
We thank the referees for their constructive comments and suggestions. 
\end{acknowledgments}

\end{document}